\documentclass[aip,graphicx,amsmath,amssymb,preprint]{revtex4-1}

\usepackage{bm}
\usepackage{graphicx}

\begin{document}


\title{Nonlinear Dynamics of Particles Excited by an Electric Curtain}


\author{Owen D. Myers}
\affiliation{Materials Science Program, University of Vermont, Burlington, Vermont 05405, USA.}
\author{Junru Wu}
\affiliation{Materials Science Program, University of Vermont, Burlington, Vermont 05405, USA.}
\affiliation{Department of Physics, University of Vermont, Burlington, Vermont 05405, USA.}
\email{Junruwu@gmail.com.}
\author{Jeffrey S. Marshall}
\affiliation{School of Engineering, University of Vermont, Burlington, Vermont 05405, USA.}



\date{\today}

\begin{abstract}
The use of the electric curtain (EC) has been proposed for manipulation and control of particles in
various applications. The EC studied in this paper is called the 2-phase EC, which consists of a
series of long parallel electrodes embedded in a thin dielectric surface. The EC is driven by an
oscillating electric potential of a sinusoidal form where the phase difference of the electric
potential between neighboring electrodes is 180 degrees. We investigate the one- and two-dimensional
nonlinear dynamics of a particle in an EC field. The form of the dimensionless equations of motion
is codimension two, where the dimensionless control parameters are the interaction amplitude ($A$)
and damping coefficient ($\beta$).  Our focus on the one-dimensional EC is primarily on a case of
fixed $\beta$ and relatively small $A$, which is characteristic of typical experimental conditions.
We study the nonlinear behaviors of the one-dimensional EC through the analysis of bifurcations of
fixed points in the Poincar\'{e} sections. We analyze these bifurcations by using Floquet theory to
determine the stability of the limit cycles associated with the fixed points in the Poincar\'{e}
sections. Some of the bifurcations lead to chaotic trajectories where we then determine the strength
of chaos in phase space by calculating the largest Lyapunov exponent.  In the study of the
two-dimensional EC we independently look at bifurcation diagrams of variations in $A$ with fixed
$\beta$ and variations in $\beta$ with fixed $A$. Under certain values of $\beta$ and $A$, we find
that no stable trajectories above the surface exists; such chaotic trajectories are described by a
chaotic (strange) attractor, for which the the largest Lyapunov exponent is found. We show the well-known stable
oscillations between two electrodes come into existence for variations in $A$ and the transitions
between several distinct regimes of stable motion for variations in $\beta$.  \end{abstract}


\maketitle 


%

\section{Introduction}
\begin{figure}[h]
\centering
\includegraphics[width= 17 cm]{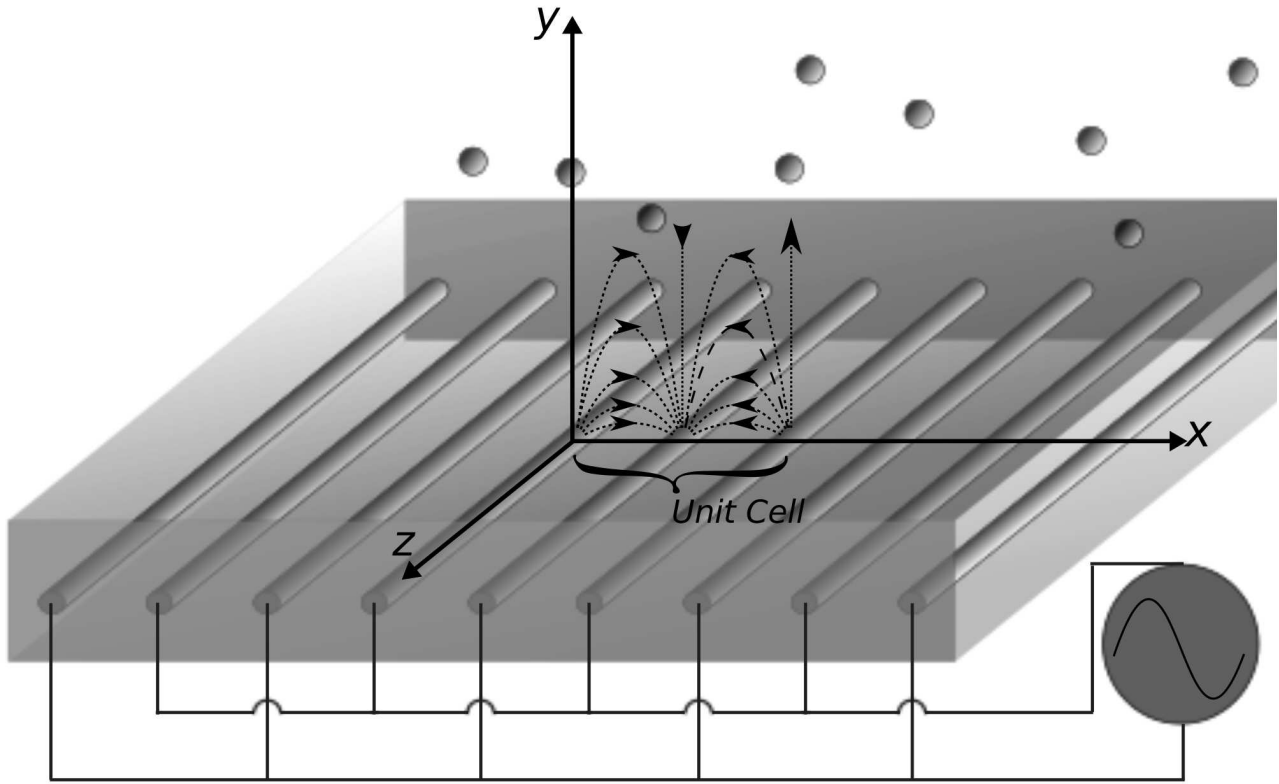}
\caption{Transparent view of a 2-phase EC with our choice of axes are super imposed. The dashed
lines represent the electric field lines in the positive $y$ plane. The 2-phase EC is periodic over the
set of electrodes marked as the "unit cell".\label{fig:ECdraw}}
\end{figure}

The electric curtain (EC) is a device consisting of a series of parallel electrodes embedded in a
dielectric surface. Alternating electric potentials are applied to these electrodes such that
neighboring electrodes have a prescribed phase difference. An illustration of a two-phase EC is
shown in fig. \ref{fig:ECdraw}. 
Different EC geometry and control parameters, which will be defined later, make
it possible to create a variety of electric fields, which can generate a wealth of physical
phenomena for charged particles. For this reason, the electric curtain has been proposed for
manipulation and control of particles in many different applications. Patented in 1974 by Senichi
Masuda, the EC was originally invented for particle control in an electrostatic powder-painting
booth \cite{patent}. Other proposed applications include separation of cells in the aqueous solution \cite{blood} and
separation of by-products from agricultural processes \cite{weiss}, transport of toner particles in
photocopying machines \cite{fredpatent}, mitigation of charged dust build-up for extra-terrestrial exploration of
dusty planets and moons \cite{dustshield}, and separation of charged particles with different charge-to-mass
ratios \cite{sepkawamoto08}.

Despite those proposed applications of ECs, few commercial applications are reported.  This may be
in part due to the fact that particle dynamics induced by an EC are complex and still not well
understood. The motion of particles in EC fields has been studied both experimentally and
computationally by a number of investigators
\cite{kawamoto,liu,jeff,dudziczlaser,fredmodes,frednolev,hemstreetveldist,3phaseveldist}. These investigations have shown a variety of
different propagating and stationary modes, including the recent report of intermittent changes of
many-body particle motion discovered by Chesnutt and Marshall \cite{chesnutt} in a discrete-element simulation
of transport on inclined ECs.

An EC configuration whose adjacent electrodes are excited by electric potentials with less than 180
degrees of phase difference will produce a traveling-wave electric field above the surface. Particle
motions in these types of fields have been found to have multifarious modes of transport, which are
commonly characterized into three categories; surfing mode, hopping mode, and curtain mode \cite{frednolev}. In
the surfing mode, particles travel synchronously with the wave front, whereas in the hopping mode
particles will stick to the surface and hop stochastically when freed by a sufficiently strong
electrostatic force or collision with another particle. In the curtain mode, high frequency electric
field oscillations force particles to be levitated above the electrode surface and travel in a
spiraling trajectory with considerably slower average forward progression speed than the
propagation velocity of the traveling wave.

In this paper we discuss the particle nonlinear motion in an electric field generated by a two-phase
EC, which has been called a “standing wave” \cite{patent,dustshield} EC. For some time it was believed that this type
of EC would have poor transport properties \cite{patent,dustshield}. However, recently both
experiments \cite{atten,confdustshield} and
numerical computations \cite{jeff} have demonstrated that 2-phase ECs can be very effective at transporting
particles, and in fact particles will exhibit two very distinct modes of transport under different
conditions.  The relative simplicity of the 2-phase EC makes it an attractive candidate for many of
the proposed applications.

A significant amount of work has been reported on ECs, but the nonlinear dynamics of the
particles within an EC field have not been studied in detail from a dynamical systems point of view.
Using simple mathematical models we have discovered very prolific behaviors of charged particles in a
two-phase oscillating EC electric field.  A detailed understanding of the rudimentary particle
motions might help us better interpret the complex phenomenon often observed in real EC
experiments. 

To simplify the mathematical presentation, we have only considered $1D$ and $2D$ motion of a single
charged particle. While this is a very simple system, it is nevertheless sufficient to exhibit a
rich variety of dynamical behavior. Using a similar method to that outlined by Masuda and Kamimura
\cite{masudaapx} for 3-phase ECs, an approximate analytical equation has been derived for the motion of a
charged particle in the 2-phase EC electric field. In general, motion of a single charged particle
in the field of the 2-phase EC is two-dimensional, provided that no initial motion in the direction along the
electrodes is introduced.  However, a special case of one-dimensional particle motion, in which
the electric field magnitude is insufficient to lift a particle initially located on the dielectric
surface of the EC, is found to generate highly interesting particle behavior.

It is worthwhile to point out that the equations of motion contain a time-dependent potential,
similar to the parametrically driven pendulum with a vertically oscillating suspension point, which
is known to have interesting dynamical properties \cite{kiminvertedpen,butikov,starrett}. Our
derived equations of motion are solved numerically and the behaviors of the particle motions induced
by the oscillating electric field are examined. We obtain limit cycles which are fixed points in the
appropriate Poincar\'{e} sections of the phase space, and bifurcations of these fixed points that lead
to chaotic motions for a range of “interaction amplitude” values (a dimensionless parameter
containing the amplitude of electrode linear charge density and the charge carried by a particle).
Linear stability for very small interaction amplitude can be analyzed using a special case of the
Mathieu equation \cite{beadhoop,mathieuapp,abramowitz,mclachlan}. As we increase the interaction
amplitude, we show a variety of interesting trajectory types in the $1D$ limit and several predominant
trajectories for the $2D$ cases.

\section{Methods}

\subsection{Model Equations}

Our model uses the so-called centerline charge approximation, in which a set of parallel electrodes
in the $x,z$ plane of the Cartesian coordinates are considered to be line filaments of infinite length
oriented along the $z$-direction \cite{masudaapx}.The $x$-$y$ plane is chosen to be orthogonal to the electrode axes
($z$-axis) as shown in fig. \ref{fig:ECdraw}, where the $x$ and $y$-axes are parallel and perpendicular to the
dielectric surface, respectively. To find the electric potential and electric field above the plane
of the electrodes produced by the set of equally spaced 2-phase electrodes, we adopt the conformal
transformation used by Masuda and Kamimura \cite{masudaapx}:  

\begin{equation}
e^{(-y+ix)2\pi/\lambda}=u+iv.
\label{eqn:conformal}
\end{equation}

\noindent
This transformation maps the $x,y$ ($y>0$) half-plane containing infinite numbers of periodic
electrodes into a
unit circle in the $u,v$ complex plane containing just two electrodes. Therefore, the electric potential
can be easily computed in the $u,v$ plane. The inverse conformal transform is then performed to map
the expression back to $x-y$ coordinates.  The detailed derivation is given in Appendix A. Based on
the derivation, the electric potential can be simply expressed as:

$$\Phi (x,y) = \frac{-Q}{4\pi\epsilon_0}\cos{\omega t}\ln{\frac{\cosh{ky}+\cos{kx}}{\cosh{ky}-\cos{kx}}}$$.

\noindent
Using $\boldsymbol{E} = -\nabla \Phi$ the electric field in the $x$ and $y$ directions is found to be

\begin{equation}
E'_x = \frac{kQ}{2\pi\epsilon_0}\sin{(kx')}\cos{(\omega t')}\frac{\cosh{(ky')}}{\cosh^2{(ky')}-\cos^2{(kx')}},
\label{eqn:Ex}
\end{equation}

\begin{equation}
E'_y = \frac{kQ}{2\pi\epsilon_0}\sinh{(ky')}\cos{(\omega t')}\frac{\cos{(kx')}}{\cosh^2{(ky')}-\cos^2{(kx')}},
\label{eqn:Ey}
\end{equation}

\noindent
where $Q$ is the linear charge density amplitude of an electrode, $m$ and $q$ are the mass and electric
charge carried by a particle, respectively, $k$ is the wave number, and $\omega$ is the angular driving
frequency of the driving electric fields.  It is noted that we use the primed variables here to save
unprimed variables for their
dimensionless counterparts, which are defined below. For the centerline charge approximation to be
valid it is required that the dielectric surface be located a minimal distance above the electrodes
\cite{masudaapx}.
However, if the surface is far from the electrodes, the cosh-term on the right-hand
side of Eq. (\ref{eqn:Ex}) and Eq. (\ref{eqn:Ey}) diminishes as $1/\cosh{y'}\sim e^{-y'}$
and consequently the electric field magnitude rapidly
approaches zero. 

The dimensionless time, horizontal and vertical positions are defined by $t=\omega t'$, $x=kx'$, and
$y=ky'$. A dimensionless interaction amplitude ($A$), gravitational acceleration ($g$) and damping
coefficient ($\beta$) are, respectively, defined by $A=\frac{k^2qQ}{4\pi\epsilon_0m\omega^2}$,
$g=g'k/\omega^2$, and $\beta=\beta'/m\omega$, where $\beta'$ is the damping coefficient, and $g'$ is
the gravitational acceleration. The dimensionless form of the differential equations governing the
particle motion are:

\begin{equation}
\frac{d^2x}{dt^2}+\beta\frac{dx}{dt}=A\sin{x}\cos{t}\frac{2\cosh{y}}{\cosh^2{y}-\cos^2{x}} ,
\label{eqn:dimlessx}
\end{equation}

\begin{equation}
\frac{d^2y}{dt^2}+\beta\frac{dy}{dt}=A\sinh{y}\cos{t}\frac{2\cos{x}}{\cosh^2{y}-\cos^2{x}}-g .
\label{eqn:dimlessy}
\end{equation}
 
\noindent
Letting the dimensionless spacing of the distance between neighboring electrodes be $\pi$, the system
is periodic over a distance $\lambda = 2\pi$ in $x$. We choose the dielectric surface to be at $y=1$,
i.e. $y' = 1/k = \lambda/2\pi$, for which the centerline charge approximation holds.

\subsection{Distinction of One and Two Dimensional Regimes}

Two distinct regimes are considered: (1) A one-dimensional regime where the particle is constrained
to roll or slide back and forth on the dielectric surface;  (2) a two-dimensional regime where the
particle moves in the x-y plane, either being levitated above the surface or bouncing off of it.
Dissipative forces are included in the model for both regimes. For the $1D$ regime, dissipative forces
may arise both from rolling resistance between the particle and the surface and from the viscous
fluid force (Stokes drag) between the particle and the surrounding air. Both of these dissipative
forces are proportional to the particle velocity \cite{ballsurf}. In the $2D$ case, the particle is assumed to
have elastic collisions with the dielectric surface, so the only dissipation is from fluid drag
force. 
The transition from motion on the surface ($1D$) to the $2D$ regime occurs when the maximum vertical
force imposed on a particle immediately above an electrode exceeds the gravitational force on the
particle. For a charged particle attached to the dielectric surface, the electric field is evaluated
at a value of $y’$ equal to the particle centroid position.  Letting the gravitational force balance
the maximum electrostatic force, the maximum value of the interaction amplitude $A$ for which the
particle remains on the surface can be obtained by setting the left-hand side of
Eq. (\ref{eqn:dimlessy}) to zero, giving

\begin{equation}
\frac{A}{g}\le\frac{cosh^2{y}-1}{2\sinh{y}}
\label{eqn:balance}
\end{equation}

\noindent
With the location of the surface at $y=1$, the critical value of this ratio for which the $1D$
approximation applies is  obtained as $A/g\le0.588$. In order to further simplify this system, we neglect 
the adhesive van der Waals force. This simple system is used to examine the nonlinear dynamics
for small variations over a range of $A$ with constant values of $\beta$.

\section{Time Maps}
We classify different trajectories by the periodicity of their limit cycles in the full phase space.
The interaction amplitude, damping, and initial conditions determine the periodicity of the particle
trajectory. We compare the temporal length of a limit cycle to the inverse of the driving frequency
of the electric field by using time maps. Time maps are used to represent the advancement of
an orbit in phase space by some amount of time. In general, a particle's position and velocity
determine its position in phase space, which can be represented as a vector function
$\boldsymbol{r}(\boldsymbol{r}_0,t)$, where $\boldsymbol{r}_0$ is the initial position in the phase
space.  We choose the time maps to be represented in the form
$\boldsymbol{f}(\boldsymbol{r}_0,t=2\pi n)$.  The time mapping gives a stroboscopic view of the
parametric function $\boldsymbol{r}$ when $n$ is a positive integer ($\mathbb{Z}^+$). For
$n\in\mathbb{Z}^+$ the time mapping highlights the relationship between the period of a trajectory
and the driving frequency. We define an operator $\boldsymbol{T}$ that maps the system forward in
time by $2\pi$, so that the function $\boldsymbol{f}(\boldsymbol{r}_0,t=2\pi n)$ may be written as
$\boldsymbol{T}^n\boldsymbol{r}$. 

\begin{figure}[h]
\centering
\includegraphics[width=8.5 cm]{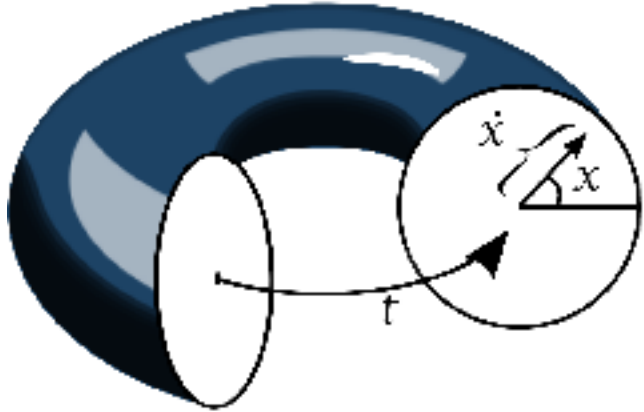}
\caption{Due to the periodicity of the EC in the $x$-direction, and the periodic fluctuations of the electric
field with time, it is useful to use toroidal geometry to describe the $1D$ EC phase space.  We define the
$t$-axis as the line at the major radius when the minor radius is zero.  The position and velocity
at a given time may then be represented as a point in the plane orthogonal to the time axis
described by an angle ($x$) and radius ($\dot{x}$). The origin of this plane is located at the
intersection with the $t$-axis. \label{fig:toroidspace}} 
\end{figure}

The $2D$ EC has five degrees of freedom $x$, $y$,
$\dot{x}$, $\dot{y}$, $t$. The periodicity of the system in $x$ and $t$ implores the use of a
toroidal phase space. We use this notion of a toroidal phase space to fashion our Poincar\'{e}
sections. We defer to the $1D$ EC in order to visualize the toroidal phase space (fig.
\ref{fig:toroidspace}). For the $2D$ EC we cannot graphically depict the full phase space but we use
the same periodic geometries to fashion Poincar\'{e} sections. A Poincar\'{e} section includes any point
where a continuous trajectory or flow transversely intersects a subspace of the space the
trajectory occupies\cite{guckenheimer}. Strobing a periodically driven system will produce
Poincar\'{e} sections of $d-1$ dimension, where $d$ is the dimension of the full phase space. This
is equivalent to observing time maps $\boldsymbol{T}^n$ acting on a point in phase space.  It
becomes apparent that this method of strobing or time mapping produces  Poincar\'{e} sections when
we look at a particle's trajectory as a function of time $t$ passing through a $x,\dot{x}$ phase
plane at a particular time $t$. The particles path is always transverse to the $x,\dot{x}$ plane and
therefore a point of intersection of the trajectory with this plane is a convenient sub space that
satisfies the criterion necessary to be a Poincar\'{e} section. It is also true that time maps
generate Poincar\'{e} sections for the $2D$ EC, but in this case the mappings represent
intersections of a trajectory with a non-planar subspace. This is still a Poincar\'{e} section
because the flow through the subspace is guarantied to be transverse to it due to the positive rate
of change of time.  

By filling a plane in phase space with an array of initial conditions, a
large number of possible trajectories can be obtained for given values of $\beta$ and $A$.
For instance, in $1D$ simulations initial conditions are prescribed in the $x-\dot{x}$ plane at $t=0$. For $2D$
simulations, initial conditions are prescribed on a planar section of the four-dimensional phase
space at $t=0$. Different initial planes are used to explore different regions of the phase space. We
define a region of phase space filled with an array of initial conditions as a \emph{block}. A group of
positions in the phase space at a given time is represented by a $2\times N$ dimensional matrix $\boldsymbol{B}$, where the first
index (2) is the dimensionality of the block and the second index ($N$) is the
number of  different initial conditions in the block. The advancement of these initial conditions in
phase space by intervals of the driving period can be represented with time maps by
$\boldsymbol{T}^n\boldsymbol{B}$. The different attractors in the system can be found by taking
$\boldsymbol{T}^n\boldsymbol{B}$ as $n\rightarrow \infty$. As discussed above, a series of time maps
can be used to generate a Poincar\'{e} sections of $\boldsymbol{B}$ in the phase space.

\section{Results}

\subsection{One Dimensional Regime}

For small $A$ and finite $\beta$, a particle on the dielectric surface will drift toward the nearest electrodes. For an
electrode located at $x=0$, an analytical solution can be obtained for particles at small distances
$|x|$ from the electrode. Using the leading-order Taylor series approximations $\sin{x}\approx x$
and $\cos{x}\approx 1$, the multiplying
factor in Eq. (\ref{eqn:dimlessx}) can be approximated as 
\begin{equation}
\frac{2\cosh{y}}{\cosh^2{y}-\cos^2{x}}\approx \frac{2\cosh{y}}{\cosh^2{y}-1} \equiv C,
\label{eqn:C}
\end{equation}

\noindent
where $C$ is a constant for $y=1$. The equations of motion then become 

\begin{equation}
\frac{d^2x}{dt^2}=ACx\cos{t}-\beta\frac{dx}{dt}.
\label{eqn:simple1D}
\end{equation}

\noindent
Under the transformation $x(t)=e^{(-\beta t/2)} u(t)$ and $t=2\theta$ it takes the form \cite{mclachlan} 

\begin{equation}
\frac{d^2u}{d\theta^2}=u(a-2q\cos{2\theta}),
\label{eqn:mathiue}
\end{equation}

\noindent
where $q=AC/8$ and $a=-\beta^2$. Equation (\ref{eqn:mathiue}) is the canonical form of the Mathieu
equation. There are infinite sets of alternating stable (periodic) and unstable solutions for
variation of the parameters $a$, $q$ \cite{morse}. Here we are only interested in the bound
solutions because unstable trajectories force the consideration of larger $|x|$ for which this
approximation breaks down. The function $u(t)$ may be expressed as a linear combination of the
cosine and sine type elliptic functions. It is well known that the stability of the elliptic
functions depends on the parameters $a$ and $q$. The stability boundary may be expressed as a
function $q(a)$. Gunderson et al.  \cite{gundersonmathieu} derive a condition for asymptotic
stability based the relationship between the two parameters.  The inequality found by Gunderson et
al. that needs to be satisfied for asymptotic stability in the EC takes the form $A<\beta^2/{2C}$.
This relationship only holds for small $a$ and $q$, but so does the analytical treatment of the EC.
We refer the reader to McLachlan\cite{mclachlan} for a thorough description of elliptic functions
and their different representations.

For larger values of $A$, particles are not necessarily confined above the
electrodes. In order to analyze this system and highlight the dimensionality of the full phase
space, it is
convenient to express Eq. (\ref{eqn:dimlessx}), with $y=1$ denoting the dielectric surface,
as a set of first-order autonomous differential equations.

\begin{eqnarray}
&& \dot{x}_1=x_2 \nonumber \\
&& \dot{x}_2=A\sin{x_1}\cos{x_3}\frac{2\cosh{1}}{\cosh^2{1}-\cos^2{x_1}}-\beta x_2 \nonumber \\
&& \dot{x}_3=1 
\label{eqn:firstorder1D}
\end{eqnarray}

\noindent
The following transformations used in Eq. \ref{eqn:dimlessx} give us Eq. \ref{eqn:firstorder1D}: $x
\rightarrow x_1$, $\dot{x}\rightarrow x_2$, and $t\rightarrow x_3$. We know from
the previous discussion of time maps that a period-$p$ fixed point in the Poincar\'{e} sections,
located at $\boldsymbol{r}_{fp}$ in the phase plane $x_1$ $x_2$ is defined by $\boldsymbol{T}^{pn}
\boldsymbol{r}_{fp}=\boldsymbol{r}_{fp}$, where $n\in\mathbb{Z}^+$.  A period-$p$ orbit (i.e limit
cycle) is one which repeats itself after $p$ driving cycles. All Poincar\'{e} sections obtained by
sequential time maps have two fixed points within the toroidal phase space (see fig.
\ref{fig:toroidspace}) located at $x_1=0$ and $x_1=\pi$. These fixed points are representative of
limit cycles of period-1 in the phase space. These two limit cycles are the only invariant sets in
the full parameter and phase space which exist for all values of $A$ and $\beta$. For definiteness,
we define $\boldsymbol{r}_{fp1}$ to be the $x_1=\pi$, $x_2=0$ fixed point in the Poincaré sections.
To distinguish between the fixed point in the Poincar\'{e} section and the corresponding limit cycle
in the phase space, we define the full invariant set composing the limit cycle as
$\{\boldsymbol{r}_{fp1}\}$, where the curly brackets denote a set.  

In general, the stability of a fixed point in the Poincar\'{e} section can
be analyzed using Floquet theory \cite{lefschetz}, details of which are provided in Appendix B. By
integrating the linearized equations of motion about a periodic orbit with periodicity $p$, 
a solution for small perturbations of the fixed-point solution lying on
the closed orbit is obtained. A similar approach is illustrated in \cite{kiminvertedpen} for Floquet theory
applied to the parametrically driven pendulum which is mathematically similar to the $1D$ EC. The
eigenvalues (Floquet stability multipliers) $\lambda_1$, $\lambda_2$ are generally complex, where a fixed point in the
Poincar\'{e} section is unstable when the magnitude of $\lambda_1$ or $\lambda_2$ is greater than unity.
In cases where
there is no analytical expression for a fixed point as a function of $A$, a polynomial curve fit
is used to estimate the fixed point as a function of $A$. 

\begin{figure}[h]
\centering
\includegraphics[width=8.5 cm]{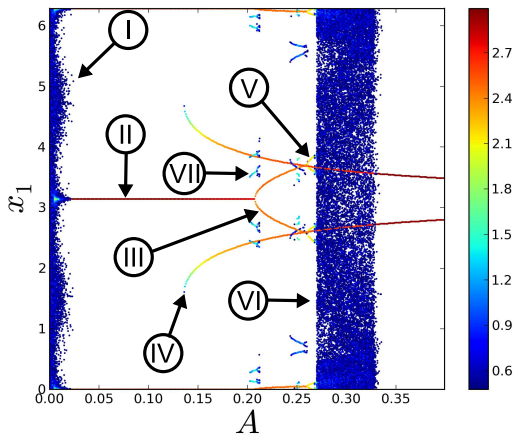}
\caption{A bifurcation diagram made by taking a two dimensional histogram of the final Poincar\'{e}
section of
1830 different initial conditions for 500 different values of $A$. $\beta$ is set to $0.1$. Roman numerals denote the
individual features focused on in the paper; Region I: With $A$ being close to zero transients take
a long time to die out. Region II: A stable fixed point above the electrode representing a sink where particles are
stationary. Region III: A period-2 orbit oscillating about the electrode. Region IV: Stable
propagating trajectories with comparatively high speeds. Region V: Four fixed points in the digram
representing two period-2 trajectories born out of a cyclic fold bifurcation. Region VI: Period
doubling leads to chaotic motion. Region VII: A period-4 fixed point in the Poincar\'{e} sections
that discontinuously appears and disappears over variations of $A$. \label{fig:labeledbif}}
\end{figure}

We initially focus attention on the $\beta = 0.1$
case because this
value is typical of a variety of realistic EC configurations. A bifurcation diagram for this
case is shown in fig. \ref{fig:labeledbif}. The points corresponding to each value of $A$ in this bifurcation
diagram are obtained by plotting the positions of a block of $1830$ different initial conditions
covering the region $0\le x_1<2\pi$ and $-1.5\le x_2 \le 1.5$, plotted after 637 time maps. This process is repeated for
different $A$ values ranging from $0$ to $0.4$, in steps of $0.0008$. In order to indicate the number of
initial conditions corresponding to each point in this diagram, a hexagonal histogram was formed
in which the $x_1$-position is discretized into 400 bins. The color bar in fig. \ref{fig:labeledbif}
represents the logarithm of the number of points in the corresponding bin.  

For very small
values of $A$, denoted by region I in fig. \ref{fig:labeledbif}, the $x_1$ position of the points in this figure are
widely dispersed. This scattering occurs because the transients are very slow to die out for
small values of $A$, and therefore the block of initial conditions has not yet reached its final
state. As we increase $A$, a single line is observed in the fig. \ref{fig:labeledbif}, denoted by region II,
representing the $\boldsymbol{r}_{fp1}$ attractor. For these values of $A$, this attractor is an asymptotically stable
fixed point in the Poincar\'{e} sections, so that initial conditions in a small neighborhood about
$\boldsymbol{r}_{fp1}$ map to $\boldsymbol{r}_{fp1}$. In 
regimes I and II, the direction of the time-averaged force on the particle points to locations of constant electrostatic
potential. This condition is similar to the well-known Bjerknes force in acoustics
\cite{cavcellbubbles}, wherein
the acoustic radiation force on a particle points to either the nodes or antinodes of a standing
acoustic field. 
The fixed point $\boldsymbol{r}_{fp1}$ remains asymptotically stable for values of A in the interval
$0<A<A_{c1}$, where $A_c1\equiv 0.20761 \pm .00001$. 

\begin{figure}[h]
\centering
\includegraphics[width=8.5 cm]{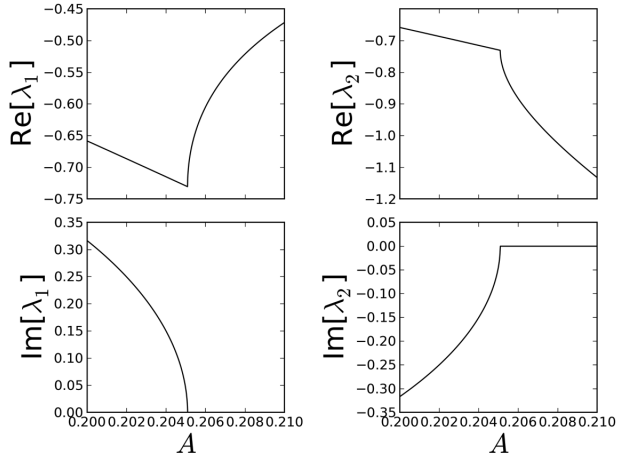}
\caption{Floquet stability multipliers of the $(x=\pi,\dot{x}=0)$ fixed point for a range of $A$
in which the first bifurcation occurs. The discontinuity is the transition from an
attracting focus to an attracting node. The supercritical flip bifurcation happens as
$\text{RE}[\lambda_2]$
becomes smaller that $-1$. \label{fig:stability}}
\end{figure}

At $A=A_{c1}$, a bifurcation of
the $\boldsymbol{r}_{fp1}$ fixed point is observed, beyond which the single line splits into two period-2 curves,
which are symmetric about the $x_1=\pi$ line as indicated in region III in fig. \ref{fig:labeledbif}. The real and
imaginary parts of the two Floquet stability multipliers, $\lambda_1$ and $\lambda_2$, for
$\boldsymbol{r}_{fp1}$ including values of $A$
close to $A_{c1}$ are shown in fig. \ref{fig:stability}. The discontinuity in fig. \ref{fig:stability}, where the imaginary parts go to zero,
represents the transition of the fixed point from an attracting focus to an attracting node. The
point at which $\lambda_2$ decreases below $-1$ in fig. \ref{fig:stability} corresponds to the bifurcation, where
$\boldsymbol{r}_{fp1}$ becomes
a period-1 saddle. Beyond this bifurcation, the fixed point becomes linearly unstable in what
is called a supercritical flip bifurcation. The instability of the period-1 saddle creates a
stable limit cycle of period-2 about $\boldsymbol{r}_{fp1}$. 
It may be surprising that the first oscillations are not period-1. This is because period-1
oscillations are not a harmonic of $\boldsymbol{r}(t)$ \cite{butikov}.
There are, however, two period-1 fixed points
apparent in fig. \ref{fig:labeledbif} seen for $A>0.13$ (one on either side of $x=\pi$) denoted by
region IV which represent 
propagating trajectories. These propagating trajectories travel once through the toroidal phase
space per driving cycle making them period-1 fixed points in fig. \ref{fig:labeledbif} but they are not periodic
oscillations because they are constantly traveling in one direction.

\begin{figure}[h]
\centering
\includegraphics[width=8.5 cm]{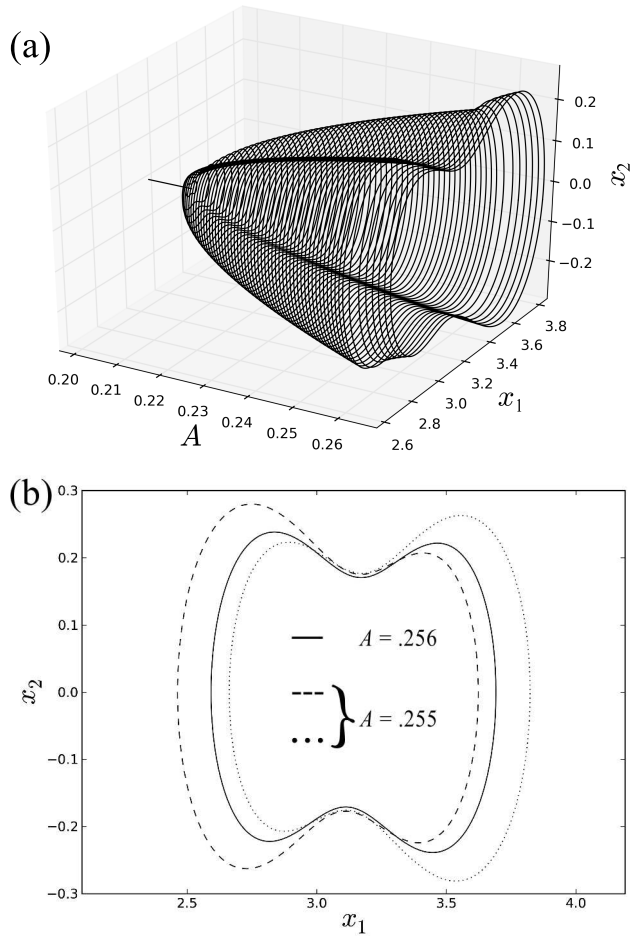}
\caption{(a) projections of the full phase space onto the $x_1$, $x_2$ axes for a range of $A$ that
includes the first two bifurcations. The first being a supercritical flip bifurcation and the second
being a cyclic fold bifurcation. (b) Solid line: Trajectory after supercritical flip but before
cyclic fold. Dashed and doted lines: The two possible trajectories after cyclic fold.\label{fig:loopsAvslim}}
\end{figure}

We call the
uppermost region III curve $\boldsymbol{r}_{fp2}$. $\boldsymbol{r}_{fp2}$ is an attracting focus for values of
$A\le 0.234$. Just as in the
first bifurcation, the attractor $\boldsymbol{r}_{fp2}$ transitions to an attracting node shortly preceding the
second bifurcation, which occurs at $A=A_{c2}$, where $A_{c2}\equiv 0.26077 \pm 0.00001$. A close-up
view showing this bifurcation is
given in fig. \ref{fig:labeledbif}b. This bifurcation is of the type known as a cyclic fold bifurcation, where
the fixed point becomes unstable and the symmetry of the trajectory corresponding to the fixed
point is broken at the bifurcation point. As $A$ is increased through $A_{c2}$ two new possible
trajectories are spontaneously created denoted by region V in fig. \ref{fig:labeledbif}.
The initial conditions determine which of the new possible trajectories a particle will settle
into. As the particle trajectory is known to be highly sensitive to small changes in initial
condition, this type of bifurcation is known to produce hysteresis for
processes in which $A$ is varied about $A_{c2}$ \cite{thompson}. An alternative viewpoint of this bifurcation
sequence is given by plotting projections along the $x_3$ ($t$) axis of the phase space onto the
$x_1$, $x_2$ plane for
some particular values of $A$ in fig. \ref{fig:loopsAvslim}a. This figure shows the fixed point $\boldsymbol{r}_{fp1}$
bifurcating into a
sequence of limit cycles, denoted in the Poincar\'{e} section by $\boldsymbol{r}_{fp2}(A)$, as $A$ is
increased past $A=A_{c1}$.
These curves are point symmetric about $(x_1=\pi,x_2=0)$ in the interval $A_{c1}<A<A_{c2}$. For $A>A_{c2}$, two
families of trajectories are
observed, one of which shifts in the positive $x$-direction and the other shifts in the negative
$x$-direction, breaking the point symmetry. We only show one of the two possible trajectories for $A>A_{c2}$ in figure \ref{fig:loopsAvslim}a for clarity.
Sample trajectories for each of these intervals are plotted on the $x_1$, $x_2$ plane in
fig. \ref{fig:loopsAvslim}b
to highlight the symmetry breaking. The first bifurcation at $A=A_{c0}$ looks geometrically similar to a
Hopf-bifurcation in fig. \ref{fig:loopsAvslim}a, but this is an artifact of the projection process along $x_3$, where in fact
the $\{\boldsymbol{r}_{fp1}\}$ set that gives rise to the $\boldsymbol{r}_{fp1}$ fixed point in fig. \ref{fig:loopsAvslim}a is
a limit cycle rather than an equilibrium point in time.   

\begin{table}[h]
\caption{Period Doubling Bifurcations}
\centering
\begin{tabular}{c c c c}
\hline\hline
$A_{cn}$ &Bifurcation & $A$ & Period of New Limit Cycle\\
\hline
$A_{c1}$ &Supercritical Flip & $0.20761 \pm 0.00001$ & $2$ \\
$A_{c2}$ &Cyclic Fold        & $0.26077 \pm 0.00001$ & $2$ \\
$A_{c3}$ &Supercritical Flip & $0.26798 \pm 0.00001$ & $4$ \\
$A_{c4}$ &Supercritical Flip & $0.26903 \pm 0.00001$ & $8$ \\
$A_{c5}$ &Supercritical Flip & $0.26928 \pm 0.00001$ & $16$  \\
$A_{c6}$ &Supercritical Flip & $0.269336\pm0.000002$ & $32$ \\ [1ex]

\hline
\end{tabular}
\label{table:bif}
\end{table}

Following this second bifurcation (region V), a series of period-doubling bifurcations occurs, with the first of these
bifurcations happening at $A=A_{c3}$, where $A_{c3}\equiv 0.26798 \pm 0.00001$.
Table \ref{table:bif} shows the values of $A$ at which each
period-doubling bifurcation occurs, the period of the limit cycle after the bifurcation and the type
of bifurcation. It can be seen in fig. \ref{fig:labeledbif} that the period doubling cascade
initial occurs in a very small range of $x$; in fact, it occurs in a very small volume of the full
phase space. The density of the higher period trajectories, even after the period-32 bifurcation,
make if extremely difficult to distinguish periodicities.
Using $A_{c3}$, $A_{c4}$, and $A_{c5}$ we calculate the Feigenbaum constant $\delta$ to be $4.2000$. Using
$A_{c4}$, $A_{c5}$, and $A_{c6}$ we calculate the Feigenbaum constant to be $4.4643$. This
progression looks as though it might lead to the universal value of $\delta\approx4.6692$
\cite{introchaos}. The Feigenbaum constant $\delta$ s be evaluated with

\begin{equation}
\delta = \lim_{n\rightarrow \infty} \frac{A_{cn-1}-A_{cn-2}}{A_{cn} - A_{cn-1}}.
\label{eqn:feigenbaum}
\end{equation}

\begin{figure}[h]
\centering
\includegraphics[width=8.5 cm]{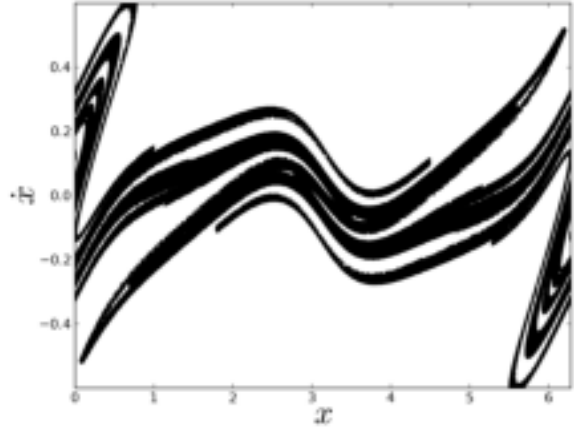}
\caption{A Poincar\'{e} section of the strange attractor found when $A=1.3$. The figure was made by
letting a trajectory approach the attractor for a long time and then plotting successive Poincar\'{e}
sections when it was assumed to be close if not in the attractor\label{fig:paperfractal}}
\end{figure}

\begin{figure}[h]
\centering
\includegraphics[width=8.5 cm]{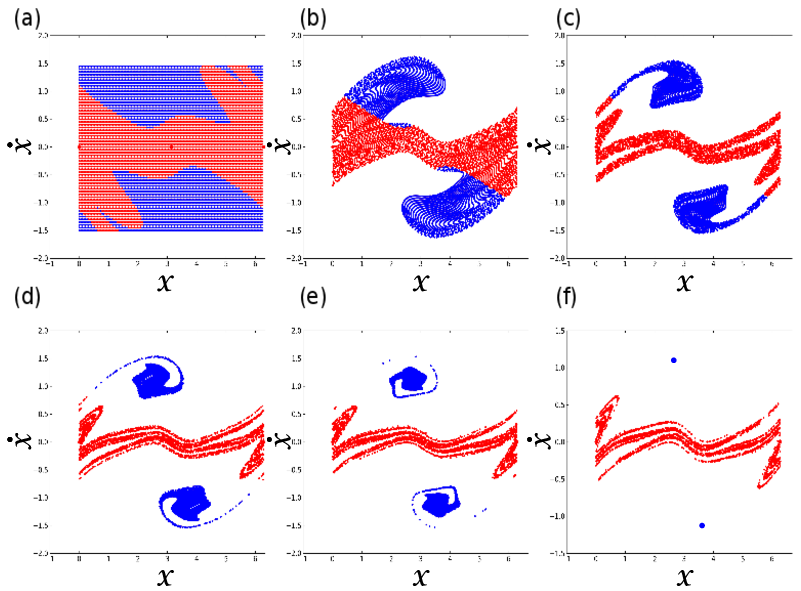}
\caption{(a) Block $\boldsymbol{B}$ of initial conditions colored red if they are in the basin of
attraction of the
strange attractor and blue if they are in the basin of the coexisting fixed points. (b)
$\boldsymbol{B}$ after time map $\boldsymbol{T}^1 $. (c) $\boldsymbol{T}^2 \boldsymbol{B}$. (d)
$\boldsymbol{T}^3 \boldsymbol{B}$. (e) $\boldsymbol{T}^4 \boldsymbol{B}$. (f) $\boldsymbol{T}^{300}
\boldsymbol{B}$. \label{fig:1Dchaosbasins}}
\end{figure}

\noindent
$A_{cn}$ are the values of $A$ for which a period-doubling bifurcation occurs, $A_{cn}$ being the
$n^{th}$ bifurcation. As $A$ is increased, the period-doubling cascade leads to a chaotic regime,
identified by the scattered points denoted by region VI in fig. \ref{fig:labeledbif}. This chaotic
state exhibits a strange attractor, which consists of a region in phase space with fractal geometry
to which particle trajectories approach as $t \rightarrow \infty$. A Poincar\'{e} section of the
strange attractor for this system is plotted in fig. \ref{fig:paperfractal}. As shown in fig.
\ref{fig:1Dchaosbasins}f, this chaotic state exists simultaneously with the two stable period-1
limit cycles (region IV), where the latter have substantial higher velocity magnitude than the
points in the chaotic state. The basins of attraction for the strange attractor and the stable limit
cycles are plotted in fig. \ref{fig:1Dchaosbasins} for $A = 0.3$, along with plots showing the
evolution of these basins in time as they approach their respective attractors. We have found the
largest Lyapunov exponent for this chaotic attractor to be $0.165\pm 0.001$. In general The largest
positive Lyapunov exponent is a way of quantifying the strength of chaos. More specifically it is
the measure of the rate with which two infinitesimally close initial conditions in phase space will
separate. We have calculated the largest Lyapunov exponent by comparing a fiducial trajectory to a
neighbor placed infinitesimally close. After a small period of time we re-orient the perturbed
trajectory along the vector for which there was maximal separation. By continuing this process for
many iterations and throwing away the transients we find the maximum separation of the perturbed
trajectory from the fiducial trajectory. With these data it is straight forward to calculate the
Lyapunov exponent\cite{thompson}.

\begin{figure}[h]
\centering
\includegraphics[width= 8.5 cm]{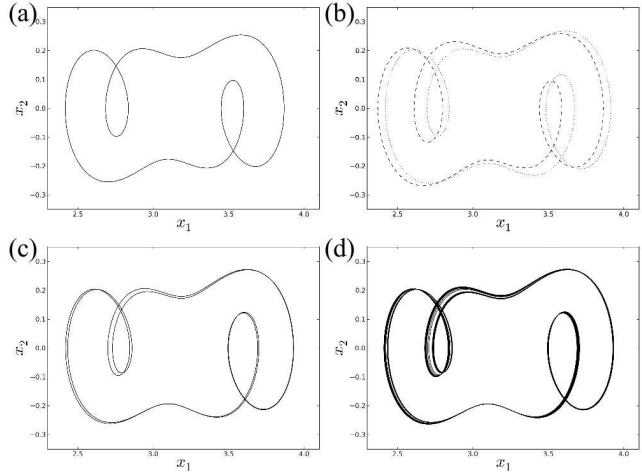}
\caption{(a) A period-4 trajectory that creates the initial region VII (\ref{fig:labeledbif}) fixed
points. (b) The two possible period-4 trajectories after a cyclic fold bifurcation distinguished by the
dashed and doted lines. (c) After the doted line in (b) undergoes a supercritical flip and its
period doubles. (d) The result of more period doubling bifurcations.\label{fig:discontloops}}
\end{figure}

The last feature we focus on in fig. \ref{fig:labeledbif} are the stable limit cycles that appear and disappear discontinuously for
small variations in $A$. We have chosen one of these limit cycles (denote as region VII in
fig. \ref{fig:labeledbif}a) to analysis the stability in order determine the nature of its
creation and disappearance. In fig. \ref{fig:labeledbif} four fixed points appear as $A$ is increased
through $0.20182\pm 0.00001$.  These fixed points correspond to a period-4 limit cycle shown in
fig. \ref{fig:discontloops} for $A=0.2086$ projected on the
$x_1,x_2$ plane. This limit cycle is point-symmetric about $(x_1=\pi,x_2=0)$ but as $A$ is increased
further the limit cycle breaks its point symmetry in a cyclic fold
bifurcation before undergoing a period doubling cascade. 

\begin{figure}[h]
\centering
\includegraphics[width=17 cm]{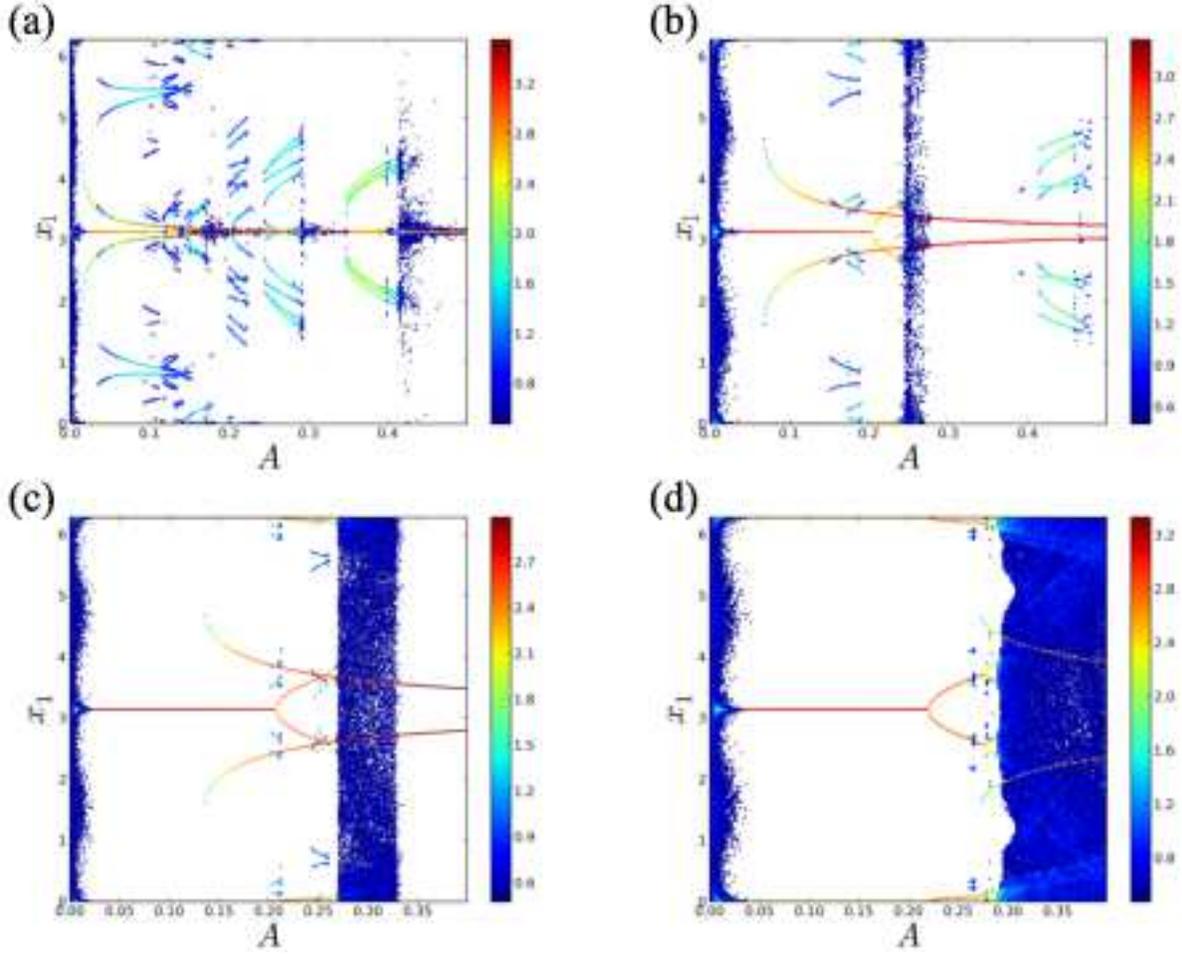}
\caption{Bifurcation diagrams computed using the same methods as in fig. \ref{fig:labeledbif}. (a) $\beta
= 0.01$. (b) $\beta = 0.05$. (c) $\beta = 0.1$. (d) $\beta = 0.2$.\label{fig:multiplebeta}}
\end{figure}

A comparative
study was performed for this system for different values of the damping coefficient $\beta$.
Bifurcation plots showing particle positions in the Poincar\'{e} sections versus the interaction
amplitude $A$ are given in fig. \ref{fig:multiplebeta} for $\beta=0.01$, $0.05$, and $0.2$, which are indicative of
systems with
very little damping, moderate damping and heavy damping, respectively. All of these plots
exhibit the region IV stable propagating period-1 limit cycles and initial stability of the $\boldsymbol{r}_{fp1}$ fixed
point at $x_1=\pi$.
In the cases of $\beta = 0.05$ and $0.2$, the $\boldsymbol{r}_{fp1}$ attractor exhibits a period-doubling bifurcation, similar
to that discussed above for the $\beta=0.1$ case, followed by a sequence of bifurcations to a chaotic
state. The $\beta=0.01$ case, by contrast, does not appear to have a clear bifurcation of the
$\boldsymbol{r}_{fp1}$
attractor. The next most evident difference between the bifurcation diagrams is the number of
trajectories that discontinuously appear and disappear as $A$ is increased. For $\beta=0.2$ we see 
fewer of these disconnected fixed points, one set being very similar to the fixed points associated
with the limit cycles shown in 
fig. \ref{fig:discontloops}. Clearly the bifurcation sequence to the chaotic attractor, the chaotic state itself,
and the higher velocity propagating trajectories are the dominate features in the larger
$\beta$ phase space.

\begin{figure}[h]
\centering
\includegraphics[width=8.5cm]{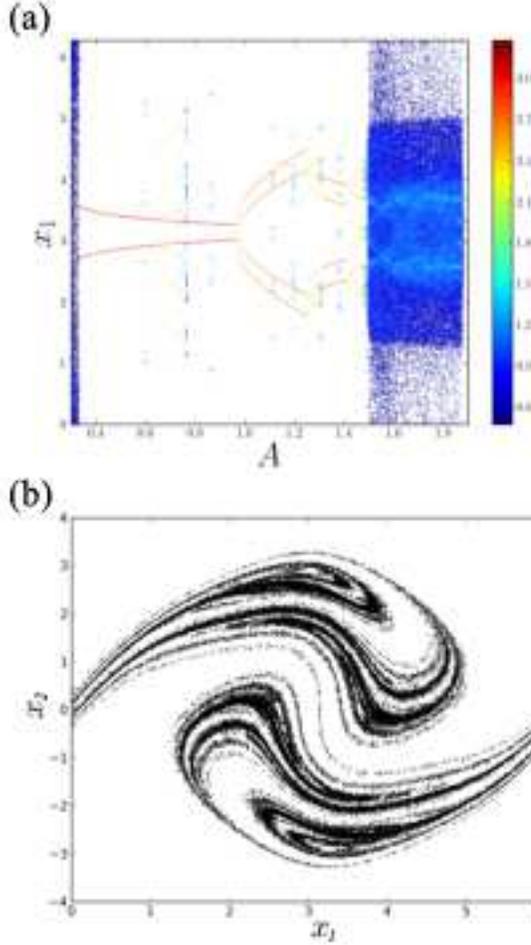}
\caption{(a) bifurcation diagram for the second range of stable
motion as well as a transition to a second chaotic regime. (b) The chaotic attractor found for
$A=1.7$.\label{fig:nextA}}
\end{figure}

We briefly continue our investigation past the first chaotic regime. We
find that as we continue to increase $A$ there are alternating arrangements of chaotic and
non-chaotic solutions. In fig. \ref{fig:nextA}a we show the
next stable regime and its transition to
chaos in a bifurcation diagram. In fig. \ref{fig:nextA}b we show
the chaotic attractor for $A=1.7$ and
the largest Lyapunov exponent for this value of the interaction amplitude is found to be
$0.187\pm .003$. 

\section{Two Dimensional Regime}
In the $2D$ EC the additional force of gravity in the equations of
motion make the system codimension 3. For the rest of this discussion we set $g=0.1$ because for certain values of $A$ and
$\beta$ it is
found to produce results on a convenient dimensionless timescale that are similar to those discussed
in the literature \cite{atten,dudziczstand,masudastand,hemstreetveldist,callemazumder,confdustshield,dustshield}.
It is worth clarifying this choice of $g$ because it may seem that
$g=0.1$ and our choice of $A$ values in the $1D$ EC section violate the
inequality in Eq. (\ref{eqn:balance}). This issue is rectified by choosing the appropriate $k$ and $\omega$
for the $1D$ EC where the ratio $k/\omega$ is
quadratic in $A$ and linear in $g$ making it possible to balance $A/g$ so that Eq. (\ref{eqn:balance})
is satisfied. We focus primarily on particle motion in the
$x,y$ plane for ease of comparison to previous and future experimental work

\begin{figure}[h]
\centering
\includegraphics[width=8.5cm]{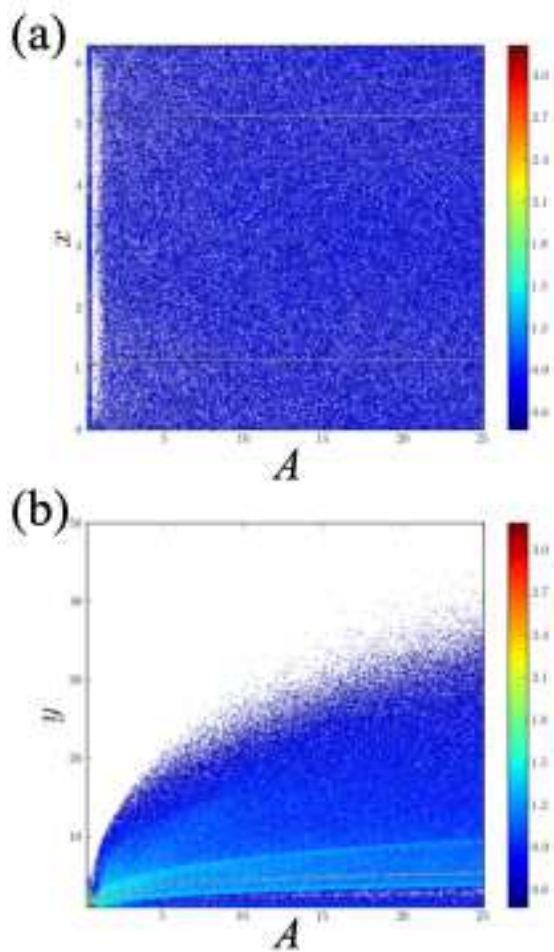}
\caption{Bifurcation diagram for $\beta = 0.1$ made with same methods as \ref{fig:labeledbif} with
projections of the final Poincar\'{e} section onto the (a) $x$ axis and (b) the $y$ axis.\label{fig:2DbifA}}
\end{figure}

To begin the study of the two dimensional regime we fix the damping at $\beta = 0.1$ and then sweep
through the interaction amplitude from $A=0.1$ to $A=25$. We use the same methods in making the
bifurcation diagrams as those used in making the $1D$ EC bifurcation diagrams. Even when no
bifurcating fixed points are found this methodology is an informative way to explore the $2D$ EC
dynamics.  In fig. \ref{fig:2DbifA}a the final Poincar\'{e} sections used to make the diagram were
projected onto the $x$ axis and in fig. \ref{fig:2DbifA}b the Poincar\'{e} sections were projected
onto the $y$ axis. In fig. \ref{fig:2DbifA}a there are two red lines trisecting the diagram in the
horizontal direction as well as a background of scattered points.  The two red lines are
asymptotically stable fixed points (attractors) in the Poincar\'{e} sections that exist for all $A$
above $A\approx 0.3$. These fixed points in the Poincar\'{e} sections are located directly
in-between the electrodes. In fig. \ref{fig:2DbifA}b there is only one red line with the background
of scattered points implying both fixed points are located at the same $y$ for a given $A$. In fig.
\ref{fig:2DbifA}a it is also clear that the location of these fixed points increases in $y$ as $A$
is increased. The two fixed points are period-$2$ limit cycles. They do not contact the surface and
they oscillate in both the $x$ and $y$ directions in an attempt to follow the curved electric field
between two electrodes (field lines can be seen in fig. \ref{fig:ECdraw}). Gravity provides a
centripetal force for the curved oscillations. The particles oscillation height depends on the force
of gravity, the time average force in $y$, and the inertial force from the particle following a
curved path.

In fig. \ref{fig:2DbifA}b the background of scattered points are seen to be restricted to a domain
of $y$ which depends on $A$. The scattered points are transient motions that take a long time to
completely die out for some initial conditions. For these parameters, as $t\rightarrow \infty$ all
initial conditions are in the basin of attraction for one of the two attractors. 

\begin{figure}[h]
\centering
\includegraphics[width=8.5cm]{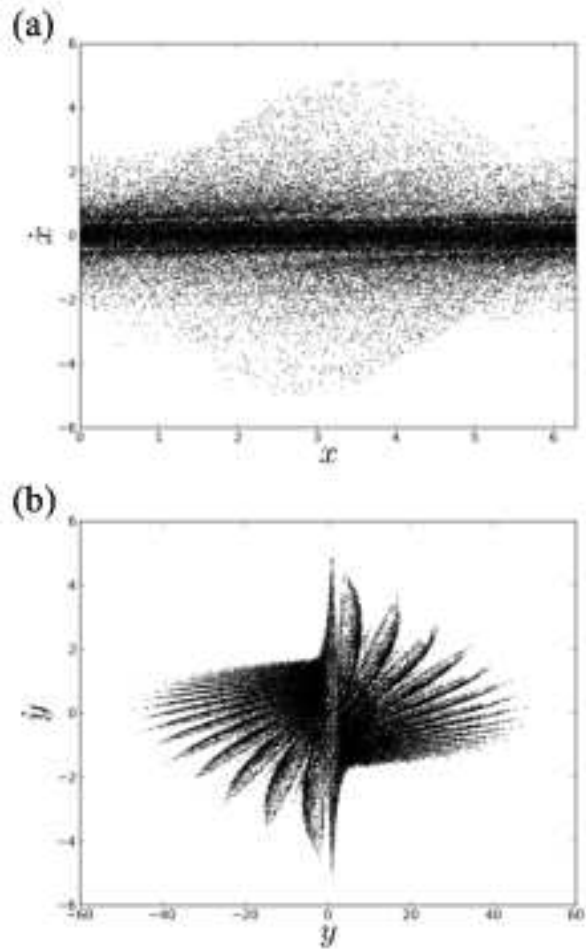}
\caption{Poincar\'{e} section of the chaotic attractor found in the $2D$ EC phase space for $A=9.0$ and
$\beta = 0.05$ plotted onto the (a) $x,\dot{x}$ plane and (b) $y,\dot{y}$ plane where we have mapped
Poincar\'{e} sections after odd numbered reflections from the surface to below the surface in order
to see the structure of the strange attractor more clearly. Made with same
method as \ref{fig:paperfractal}.\label{fig:2Dchaos}}
\end{figure}

For $\beta=0.05$ a strange attractor exists, shown for $A=9.0$ projected onto the $x$ and $y$ phase
planes in fig. \ref{fig:2Dchaos} respectively. If a stable limit cycle can not exist because $\beta$
is too small or $A$ is too large then the only type of motion found is the strange attractor in fig.
\ref{fig:2Dchaos} or a qualitatively similar strange attractor.  The strange attractor is robust for
variations in $A$ as it only grows (shrinks) in the $y$, $\dot{y}$ directions when $A$ is increased
(decreased).  The largest Lyapunov exponent for $A=9.0$ is $0.134\pm 0.003$.  

\begin{figure}[h]
\centering
\includegraphics[width=17cm]{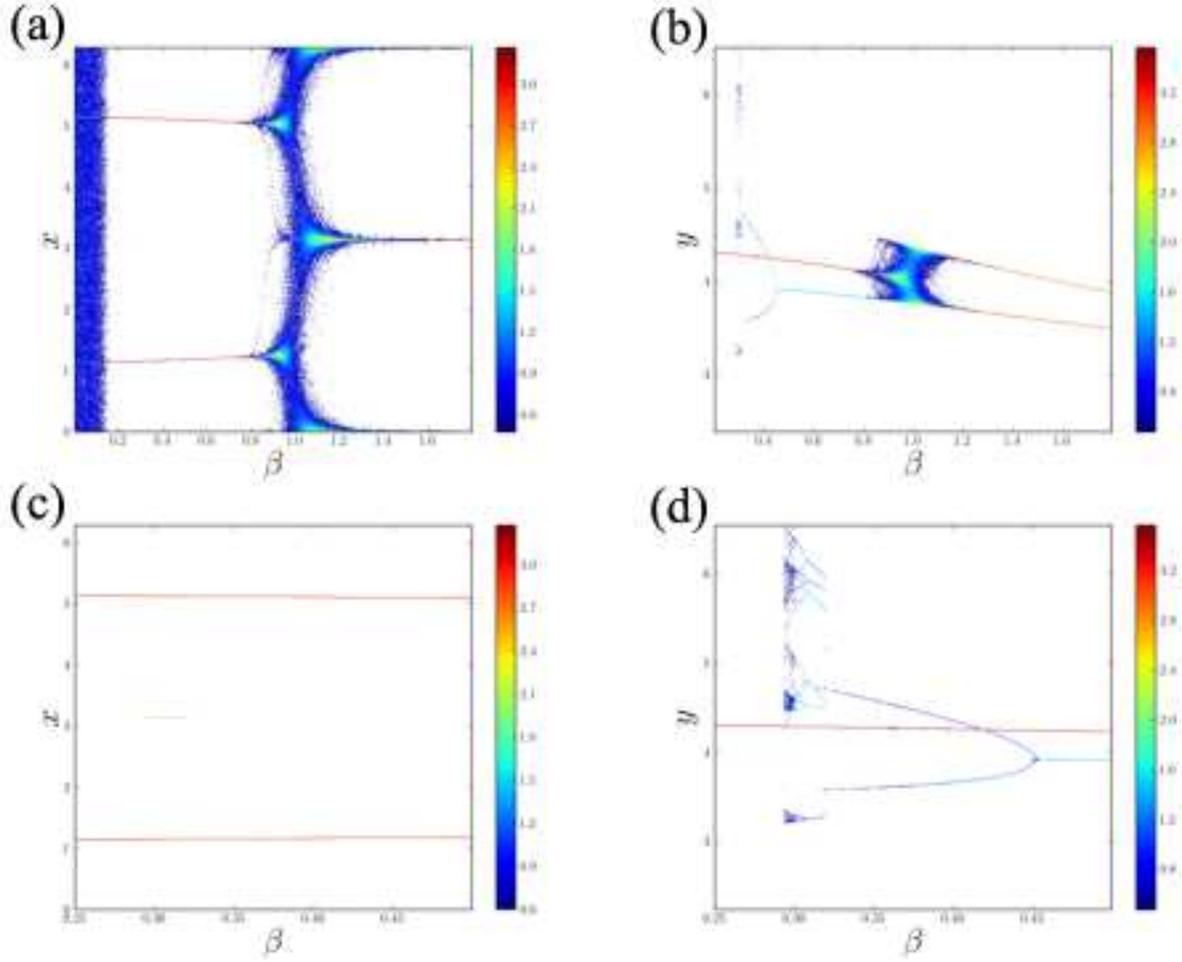}
\caption{Bifurcation diagram for variation in $\beta$ with $A=9.0$. Final Poincar\'{e} sections
projected on to (a) $x$ axis $0.0<\beta<1.8$, (b) $y$ axis $0.2<\beta<1.8$. In (c) (projections onto
$x$ axis) and (d) (projections onto $y$ axis) we show
bifurcation diagrams for $0.25<\beta<5$ to highlight the reverse bifurcation sequence. \label{fig:2DbifB}}
\end{figure}

Even though independent variations of the damping are much more difficult to achieve experimentally
we find that these variations produce slightly more interesting bifurcation diagrams. In fig.
\ref{fig:2DbifB} we show $x$ and $y$ "bifurcation" diagrams for $A=9.0$. In fig. \ref{fig:2DbifB}a,b
we show results for $\beta<2.0$ and in fig. \ref{fig:2DbifB}c,d we show results for
$0.25<\beta<2.5$. Figure \ref{fig:2DbifB}a starts with a background of scattered points coexisting
with two fixed points between the electrodes which is maintained for $0.0<\beta<0.15$.  For
$\beta>0.9$ the chaotic trajectory (fig. \ref{fig:2Dchaos}) drops out and only the stable fixed
points exist until $\beta\approx 0.9$. For $0.9<\beta<1.1$ the fixed points quickly loose their
stability to a brief period of what misleadingly appears as another chaotic regime (explanation
below). For $\beta>1.1$ two new stable fixed points are seen located in $x$ at the position of the
electrodes. In fig. \ref{fig:2DbifB}b we only show the bifurcation diagram for $0.2<\beta<2.0$ in
order to show the most important features. In this diagram a red line, representing the two initial
stable fixed points at the same value of $y$, becomes unstable at $\beta\approx 0.9$. We then see
more scattered points which are more localized in $y$ than in $x$. Then two new fixed points appear
out of the unstable region for $\beta>1.1$.  The two fixed points in $y$ after $\beta\approx 1.1$
show that there are now four fixed points in the $x,y$ plane. The two new fixed points located over
each electrode represent a limit cycle that oscillates almost entirely in $y$ with a small $x$
component oscillation as well. These are stable limit cycles of particles attesting to follow the
field lines near the electrodes.  The most notable feature of fig. \ref{fig:2DbifB}b, however, is
the apparent reverse bifurcation cascade beginning at $\beta = .29$ which we discuss in more detail
at the end of this section. 

The fixed points seen for $\beta<0.9$ in fig. \ref{fig:2DbifB} are the Poincar\'{e} sections of the
same type of stable limit cycle discussed for fig. \ref{fig:2DbifA}. The instabilities arise at
$\beta=0.9$ because the particle can no longer maintain large oscillations that following the sharp
curvature of the field directly between the electrodes. As the $x$ component of their oscillations
begins to damp out the time average force in $x$, which points towards the direction of constant
potential (directly over the electrodes) begins to weaken the stability of the fixed points between
the electrodes. The scattered points we see in fig. \ref{fig:2DbifB}a for $\beta = 1.0$ are the
result of the competing stabilities of the fixed points between the electrodes and the fixed points
above the electrodes. This competition creates a regime where particles will oscillate about any
point in $x$ but only about specific values of $y$ in an attempt to follow the local field
oscillation. These fixed points in the Poincar\'{e} sections may be described as being asymptoticly
stable in $y$ and Lyapunov stable in $x$. Meaning that a particle oscillating at a some $x$ will
continue to oscillate about that location unless there is a small perturbation in $x$ after which it
will oscillate about its new perturbed location. By Contrast small perturbations in $y$ of an
oscillations will push the particle out of equilibrium and it will experience a restoring force back
to its original position. The resulting attractor appears as a line in the Poincar\'{e} sections. As
we continue to increase $\beta$ past $1.0$ the time average force in $x$ begins to dominate particle
behavior.  The fixed points that become most clear for $\beta>1.1$ represents stable oscillations in
$y$ above an electrode. The particles settle at a height above the surface for which the time
averaged force in $y$ and $g$ are balanced. 

Figure \ref{fig:2DbifB}c shows the only actual bifurcation sequence found for the $2D$ EC, but from
fig. \ref{fig:2DbifB}d we see that this is actually behavior for particles constrained to $1D$
motion in $y$ directly above electrodes.  For this range of $\beta$ the time average force in $x$ is
not prevalent enough to prevent stable oscillations between the electrodes, as those are apparent in
fig. \ref{fig:2DbifB}c,d as well, but it is substantial enough to constrain the oscillations of
particles near electrodes to motion directly above them. It is interesting to note the existence of
such behavior in this model. We do not investigate it in any detail because observation of any such
dynamical behavior in a stable reproducible way would be extremely difficult experimentally and has
never been discussed in the literature.

\section{Conclusion}
We have studied the dynamics of a single particle in a 2-phase EC. We have separately considered the
case of particle motion constrained to the surface of an EC for small interaction amplitude ($1D$ EC)
and the case of particle motion above the surface when the interaction amplitude is sufficiently
large to lift the particle off of the surface ($2D$ EC). We find a wide variety of possible stable
limit cycles with different periodicities in the $1D$ EC and show the bifurcations of fixed points in
the Poincar\'{e} sections for variations in $A$.  For limit cycles in the $1D$ EC we calculate Floquet
stability multipliers in order to analyze the transitions found in the bifurcation diagrams. We show
that in the $1D$ system it is possible to have different trajectories coexisting for the same values
of $A$ and $\beta$. In particular we find a chaotic trajectory that coexists with two asymptoticly
stable propagating trajectories having $\pm\dot{x}$ velocities respectively. We also find that in
the $1D$ EC a transition of a limit cycle in or out of stability can happen discontinuously over small
variations in $A$. In general the number of these discontinuous trajectories is greater for smaller
values of $\beta$. We find a very different picture in the $2D$ EC. Starting with the well known
stable oscillations between two electrodes\cite{dustshield,confdustshield,dudziczstand,atten,jeff} we show how the height of this limit cycle depends
on the interaction amplitude. In an interesting transition occurring for increasing values of
$\beta$, the limit cycles looses its asymptotic stability in $x$ and a line attractor in the Poincar\'{e}
sections briefly describe particle behavior. Further increase of $\beta$ leads to the asymptotic
stability of $1D$ limit cycles in $y$ located directly above the electrodes.  In both the $2D$ and the
$1D$ models, we find chaotic motion of particles for particular parameter values. However, the
transition from stable limit cycles to chaos is fundamentally different in the two models. In the $1D$ EC
a chaotic trajectory comes out of a period doubling cascade. In the $2D$ EC, the surface interferes
with what would otherwise be a stable limit cycle and the result is chaotic motion.

By showing the general structure of particle dynamics for various values of the dimensionless
parameters $A$ and $\beta$, we develop a better understanding of how to induce and avoid certain
types of particle behavior. We believe that the sensitivity of limit cycle periodicity on the
interaction amplitude may be useful in particle sorting/separation applications. Particles with
different charge-to-mass-ratios will fall into different regimes of the parameter space. Particles
with charge-to-mass-ratios that place them in a regime of non-propagating stable motion (in either
the $1D$ or $2D$ EC) will be trapped by the EC. For example, a particle in the $1D$ EC could be
trapped in the $\beta=0.1$ and $0.0<A<0.26$ regime. The EC may be configured so that particles of a
slightly different charge-to-mass-ratio exhibit chaotic behavior by falling into the $\beta=.1$,
$0.27<A<0.33$ regime resulting in slow mitigation over time. 

The 2-phase EC is an attractive candidate for many particle manipulation and control applications,
especially for dust and particle mitigation\cite{dustshield}. We have shown that there exists
two dominate regimes of particle behavior in the $2D$ EC; one of which is stable oscillations, the
other is chaotic. Therefore for EC mitigation applications it is necessary to understand the role of
the surface as it is the primary instigator of non-stable particle trajectories.  Mitigation
efficiencies could be improved by an EC design in which most of the particles fall into chaotic
motion. This can be done by tuning the EC parameters so that charge-to-mass-ratios of interest will
exist in an area of the parameter space for which no stable motion is possible. For many real
applications, we certainly would face a system of multiple particles.  The analysis provided in the
current paper is useful for a system of dilute particle density, for which the separation distance
between particles is large enough that each particle's motion can be considered independent.  While
these results cannot be directly carried over to the case of more concentrated particle flows, they
nevertheless provide and important step in understanding the dynamics of the 2-phase EC.  

\begin{acknowledgments}
This work was supported by NASA Space Grant Consortium under grant numbers NNX10AK67H and
NNX08AZ07A. We would also like to acknowledge Chris Danforth for his insights, suggestions, and
guidance.
\end{acknowledgments}

\appendix
\section{Solving for the Electric Potential of a 2-Phase Electric Curtain}
We can greatly simplify the problem of solving for the potential of an infinite series of parallel
electrodes of infinite long length by rotating our coordinate system onto the complex plane and then
making an appropriate conformal transformation. Here we use a similar procedure to that used by
Masuda and Kamimura \cite{confdustshield} to solve for the electric potential and field of the
2-phase EC.  We start by defining the plane of the electrodes to be the $x,z$ plane. The electrodes
are infinitely long running parallel to the $z$-axis and are spaced evenly along the $x$-axis. We
label . The $y$-axis is perpendicular to the plane of the electrodes. We define the wavelength to be
twice the distance between adjacent electrodes and label them from $1,2,...,n$ respectively.  The electric
potential applied to the adjacent electrodes are 180 degrees out of phase so we express the charge
per unit length on each electrode as 

\begin{equation}
Q_n = Q\cos{\omega t - (n-1)\pi}
\label{eqn:Qn}
\end{equation}

\noindent
where $n$ denotes the electrode
of interest; i.e. $n=1,2,3,...\infty$. We choose to solve for the potential in the complex plane
$u,iv$. The mapping is
accomplished through the conformal transformation: 

\begin{equation}
e^{(y+ix)2\pi/\lambda}= u + iv
\label{eqn:Aconformal}
\end{equation}

\noindent
introduced by Masuda and Kamimura \cite{confdustshield}.
This transformation maps an infinite set of electrodes onto a unit circle containing
only two points. Electrodes $1,3,5...$ map to the same location in the $u,v$ plane as electrode 1.
It is also true that the electrodes $4,6,8...$ map to electrode 2 in the $u,v$ plane and indeed all electrodes
on $x$ map to the two points on the $u$-axis.  We can now solve for the electric potential in the
$u,v$ plane and transform our result back onto the $x,y$ plane.  Since the electrodes are infinitely long,
we may express a two dimensional electric potential produced by the two neighboring electrodes as,

\begin{equation}
\Phi (u,v)=\sum_{n=1}^{2}\frac{-Q_n}{2\pi\epsilon_0}\ln{r_n},
\end{equation}

\noindent
where $n$ denotes the $n^{th}$ electrode, $Q_n$ is given by Eq. (\ref{eqn:Qn}) and $r_n$ is the
distance between the point of interest $(u,v)$ of the field coordinates and the source coordinates
$(u',v')$ of the $n^{th}$ electrode: 

\begin{equation}
r_n=\sqrt{(u-u'_n)^2+(v-v'_n)^2}
\label{eqn:rn}
\end{equation}

Using Eq. (\ref{eqn:rn}) we can express $u$ and $v$ in terms of $x$ and $y$ and find the
corresponding electric
potential in the $x,y$ plane: $u = \operatorname{Re}\lbrack e^{k(y+ix)} \rbrack$, 
$v = \operatorname{Im}\lbrack e^{k(y+ix)} \rbrack$.
We find that $r_n = \sqrt{2e^{ky}(\cosh{ky}-\cos{k(x-x'_n)})}$.  The potential can now be expressed in terms of
only $x$, and $y$ as

\begin{equation}
\Phi (x,y)= \sum_{n=1}^{2}\frac{-Q}{4\pi\epsilon_0}\cos{(\omega t - (n-1)\pi))} \{
    \ln{2}+ky+\ln{(\cosh{ky} - \cos{k(x-x'_n)})} \}.
\end{equation}

\noindent
We can simplify this equation further by substituting in $x'_n = \frac{(n-1)\lambda}{2}$ and
noticing that $\sum_{n-1}^{2}\cos{(\omega t - (n-1)\pi)}=0$ and only
$\ln{(\cosh{ky}-\cos{k(x-x'_n)})}$ in \{ \} survives, thus

\begin{equation}
\Phi (x,y) = \frac{-Q}{4\pi\epsilon_0}\cos{\omega t}\ln{\frac{\cosh{ky}+\cos{kx}}{\cosh{ky}-\cos{kx}}}
\label{eqn:appendix_potential}
\end{equation}

\noindent
Since $\vec{E}= -\nabla \Phi$, we obtained Eq. (\ref{eqn:Ex}) and Eq. (\ref{eqn:Ex}).

\section{Floquet Stability Analysis}
Here we explicitly show the methods for stability analysis of $1D$ EC limit cycles. In order to
quantify the stability of a period-$p$ limit cycle we use Floquet theory \cite{guckenheimer}. For a
period-$p$ fixed point $\boldsymbol{r}_{fp}$ (for $\boldsymbol{r}_{fp}=(x_{1fp},x_{2fp})$) under
$\boldsymbol{T}^p$ (Poincar\'{e} map of period-$p$) the stability of $r_{fp}$ for $\boldsymbol{T}^p$
expresses the stability of the limit cycle period-$p$ $\{\boldsymbol{r}_{fp}\}$. For the rest of
this discussion we work with the non-autonomous expressions of the equations of motion and therefore
we will refer to $\{\boldsymbol{r}_{fp}\}$ as $\boldsymbol{r}_{fp}(t)$ from here on. We now
introduce the linearized-map matrix $\boldsymbol{DT}^p$, which acts on the period-$p$ fixed point
$\boldsymbol{r}_{fp}$ in the expected way i.e. $\boldsymbol{DT}^p
\boldsymbol{r}_{fp}=\boldsymbol{r}_{fp}$. The stability of $\boldsymbol{r}_{fp}$ is then found by
finding the eigenvalues of $\boldsymbol{DT}^p$ where $\boldsymbol{DT}^p$ is found by integrating the
linearized differential equations of motion for a trajectory close to $\boldsymbol{r}_{fp} (t)$. We
find a close perturbation of $\boldsymbol{r}_{fp} (t)$, which we call $\alpha (t) \equiv
(x_{1\alpha},x_{2\alpha})$, by using an infinitesimal perturbation to $\boldsymbol{r}_{fp}$ as the
initial conditions to solve for the motion $\alpha (t)$. The time dependent equations of motion for
$\alpha (t)$ may be expressed as

\begin{equation}
\binom{\dot{x}_{1\alpha}}{\dot{x}_{2\alpha}}=\boldsymbol{J}(t)\binom{x_{1\alpha}}{x_{2\alpha}},
\label{eqn:linearizedEOMwithJ}
\end{equation}

\noindent
where $\boldsymbol{J}(t)$ is
the Jacobean matrix for $\boldsymbol{r}_{fp} (t)$ and is found to be: 

\begin{equation}
    \boldsymbol{J}(t)=
    \begin{pmatrix} 
    0 & 1 \\
    \frac{\cos{t}\cos{x_{1fp}}\cosh{1}(\cosh{2} + \cos{2x_{1fp}}-2)}{(\cos^2{x_{1fp}}-\cosh^2{1})^2} & -\beta
    \end{pmatrix} .
\end{equation}

In general the solution to $\alpha (t)$ may be expressed as 

\begin{equation}
\binom{x_{1 \alpha}(t)}{x_{2\alpha}(t)}=x_{1\alpha}(0)\boldsymbol{w}^1(t)+x_{2\alpha}(0)\boldsymbol{w}^2(t),
\label{eqn:linearcombow}
\end{equation}

\noindent
where $\boldsymbol{w}^1 (t)$ and $\boldsymbol{w}^2 (t)$ are the two linearly independent solutions which can be written
together as  

\begin{equation}
    \boldsymbol{W}(t)=
    \begin{pmatrix}
    w^1_1 (t) &  w^1_2 (t)\\
    w^2_1 (t) &  w^2_2 (t)\\
    \end{pmatrix}.
\end{equation}

\noindent
This is called the solution matrix.  We can now express Eq. (\ref{eqn:linearcombow}) as 

\begin{equation}
\binom{x_{1\alpha}(t)}{x_{2\alpha}(t)}=\boldsymbol{W}(t)\binom{x_{1\alpha}(0)}{x_{2\alpha}(0)},
\label{eqn:xteqwx0}
\end{equation}

\noindent
where $\boldsymbol{W}(0)$ is necessarily the identity matrix. Now we see that $\boldsymbol{W}(t)$ is the
linearized flow $\boldsymbol{DT}^t$ and by
substituting Eq. (\ref{eqn:xteqwx0}) into Eq. (\ref{eqn:linearizedEOMwithJ}) we obtain the initial value problem 

\begin{equation}
\dot{\boldsymbol{W}}(t)=\boldsymbol{J}(t)\boldsymbol{W}(t),
\label{eqn:dotwjw}
\end{equation}

\noindent
which we can solve for $\boldsymbol{W}(t)$. For the desired period-$p$, namely the one associated with
$\boldsymbol{r}_{fp} (t)$, we numerically integrate Eq. (\ref{eqn:dotwjw})
from $t=0$ to $2\pi p$,
giving us the linearized map matrix $\boldsymbol{DT}^p$. The eigenvalues of this matrix are the Floquet stability
multipliers of $\boldsymbol{r}_{fp} (t)$ under the appropriate Poincar\'{e} map.

\end{document}